\documentclass[12pt]{article}
\usepackage[a4paper, total={17cm, 25cm}]{geometry}
\usepackage[utf8]{inputenc}
\usepackage{authblk}
\usepackage{graphicx}
\usepackage{xcolor}
\usepackage{amsmath}
\usepackage{amssymb}
\usepackage{amsfonts}
\usepackage{dsfont}
\usepackage{diagbox}
\usepackage{upgreek}
\usepackage{hyperref}

\title{\bf Single-shot self-supervised particle tracking}

\author[1]{Benjamin Midtvedt}
\author[1]{Jes\'{u}s Pineda}
\author[1]{Fredrik Sk{\"a}rberg}
\author[2]{Erik Ols\'{e}n}
\author[1]{Harshith Bachimanchi}
\author[3]{Emelie Wes{\'e}n}
\author[3]{Elin K. Esbjörner}
\author[4]{Erik Selander}
\author[2]{Fredrik H{\"o}{\"o}k}
\author[1]{Daniel Midtvedt}
\author[1]{Giovanni Volpe}

\affil[1]{{\small Department of Physics, University of Gothenburg, Gothenburg, Sweden}}
\affil[2]{{\small Department of Physics, Chalmers University of Technology, Gothenburg, Sweden}}
\affil[3]{{\small Department of Biology and Biological Engineering, Chalmers University of Technology, Gothenburg, Sweden}}
\affil[4]{{\small Department of Marine Sciences, University of Gothenburg, Sweden}}

\begin{document}

\maketitle

\begin{abstract}
Particle tracking is a fundamental task in digital microscopy. 
Recently, machine-learning approaches have made great strides in overcoming the limitations of more classical approaches. 
The training of state-of-the-art machine-learning methods almost universally relies on either vast amounts of labeled experimental data or the ability to numerically simulate realistic datasets.
However, the data produced by experiments are often challenging to label and cannot be easily reproduced numerically.
Here, we propose a novel deep-learning method, named LodeSTAR (Low-shot deep Symmetric Tracking And Regression), that learns to tracks objects with sub-pixel accuracy from a single unlabeled experimental image. This is made possible by exploiting the inherent roto-translational symmetries of the data.
We demonstrate that LodeSTAR outperforms traditional methods in terms of accuracy.
Furthermore, we analyze challenging experimental data containing densely packed cells or noisy backgrounds. 
We also exploit additional symmetries to extend the measurable particle properties to the particle's vertical position by propagating the signal in Fourier space and its polarizability by scaling the signal strength. 
Thanks to the ability to train deep-learning models with a single unlabeled image, LodeSTAR can accelerate the development of high-quality microscopic analysis pipelines for engineering, biology, and medicine.
\end{abstract}

The study of biological systems often requires tracking individual objects, from microorganisms to biomolecules \cite{Manzo2015,Midtvedt2021Quantitative}.
For example, individual particle tracking enables the study of the life cycle and proliferation of single microorganisms \cite{Krishnamurthy2020} as well as the mobility of inter- and intra-cellular particles \cite{Geerts1987,Zahid2018}. 
On the molecular level, high-contrast imaging of biological processes with unprecedented spatio-temporal resolution has been made possible by single-molecule fluorescence and super-resolution microscopy \cite{Liu2017}. 
However, like many tasks in computer vision, object tracking is surprisingly difficult and it is now often the limiting factor in the analysis of microscopic images.

Recently, deep learning has been successfully employed to improve object tracking, outperforming more standard methods, e.g., to track particles in noisy images \cite{Newby2018,Helgadottir2019}, to push the limits of super-resolution fluorescence \cite{Speiser2021}, and to quantitatively characterize sub-wavelength particles \cite{Midtvedt2021}.
The most widely used deep-learning methods use supervised learning, where a neural network is trained to solve a particular problem using large amounts of high-quality training data consisting of input data and corresponding expected results (ground truth).
Obtaining these datasets represents the effective bottleneck in the application of deep learning to particle tracking \cite{Midtvedt2021Quantitative}.
While public datasets are available, they are often inadequate to represent the idiosyncrasies of any specific experimental sample and setup.
Thus, acquiring the needed experimental datasets in house is often the only viable option, but it comes with its own burdens in terms of time and effort.
%, and may discourage further developments of the experimental setup that would invalidate the collected data.
As a consequence, most deep-learning methods for particle tracking rely on synthetic data \cite{Newby2018, Helgadottir2019, Speiser2021, Midtvedt2021}. However, accurate synthetic replication of experimental data is very challenging, even for relatively simple transmission microscopes \cite{Midtvedt2021Quantitative}.

Even if a sufficiently large dataset can be collected, determining the corresponding ground truth with sufficient accuracy can be even more challenging. 
Human-derived annotation of experimental data is highly labor-intensive and prone to inconsistencies \cite{Ulman2017}, especially when dealing with high-noise data or when sub-pixel precision is required. 
Automatically-generated annotations can be employed by mimicking an existing method, but this relies on a prior ability to algorithmically analyze the data --- and the resuting deep-learning method is unlikely to outperform the original method.
Finally, numerically-generated data naturally come with the exact ground truth used to generate them, but their applicability is limited to experimental setups that can be recreated numerically, which is only possible in some cases.

Here, we tackle these issues by developing a novel deep-learning approach, named LodeSTAR (\textbf{Lo}w-shot \textbf{de}ep \textbf{S}ymmetric \textbf{T}racking \textbf{A}nd \textbf{R}egression), that exploits the inherent symmetries of particle tracking to enable training on extremely small datasets (down to a single particle image) without ground truth. 
We achieve this by merging two recent advances in deep learning. 
First, we employ \emph{geometric deep learning} to dramatically reduce the amount of required training data by exploiting the symmetries inherent in particle tracking, e.g., the fact that a roto-translation of the input particle image relates to an equal roto-translation of the expected result.
Second, we employ \emph{self-distillation} to part with the need to know the ground truth by generating a pseudo-result, e.g., from a prediction of the position of an object in an image, we can generate a ground truth for a translated view of the same object by considering the translational symmetry of the problem.
We demonstrate that a \emph{single} training image is sufficient for training LodeSTAR to outperform standard methods in terms of accuracy, while simultaneously providing robust tracking in complex experimental conditions such as densely-packed or noisy images.
Furthermore, beyond in-plane particle tracking, we demonstrate that it is possible to exploit additional symmetries to measure other particle properties for which no widespread standard methods are available, e.g., the particle vertical position exploiting the propagation symmetry of the image in Fourier space and the particle polarizability by exploiting the scaling symmetry of the signal strength of the image.

\section*{Results}

\subsection*{Geometric self-distillation}

\begin{figure}
    \centering
    \includegraphics[width=\columnwidth]{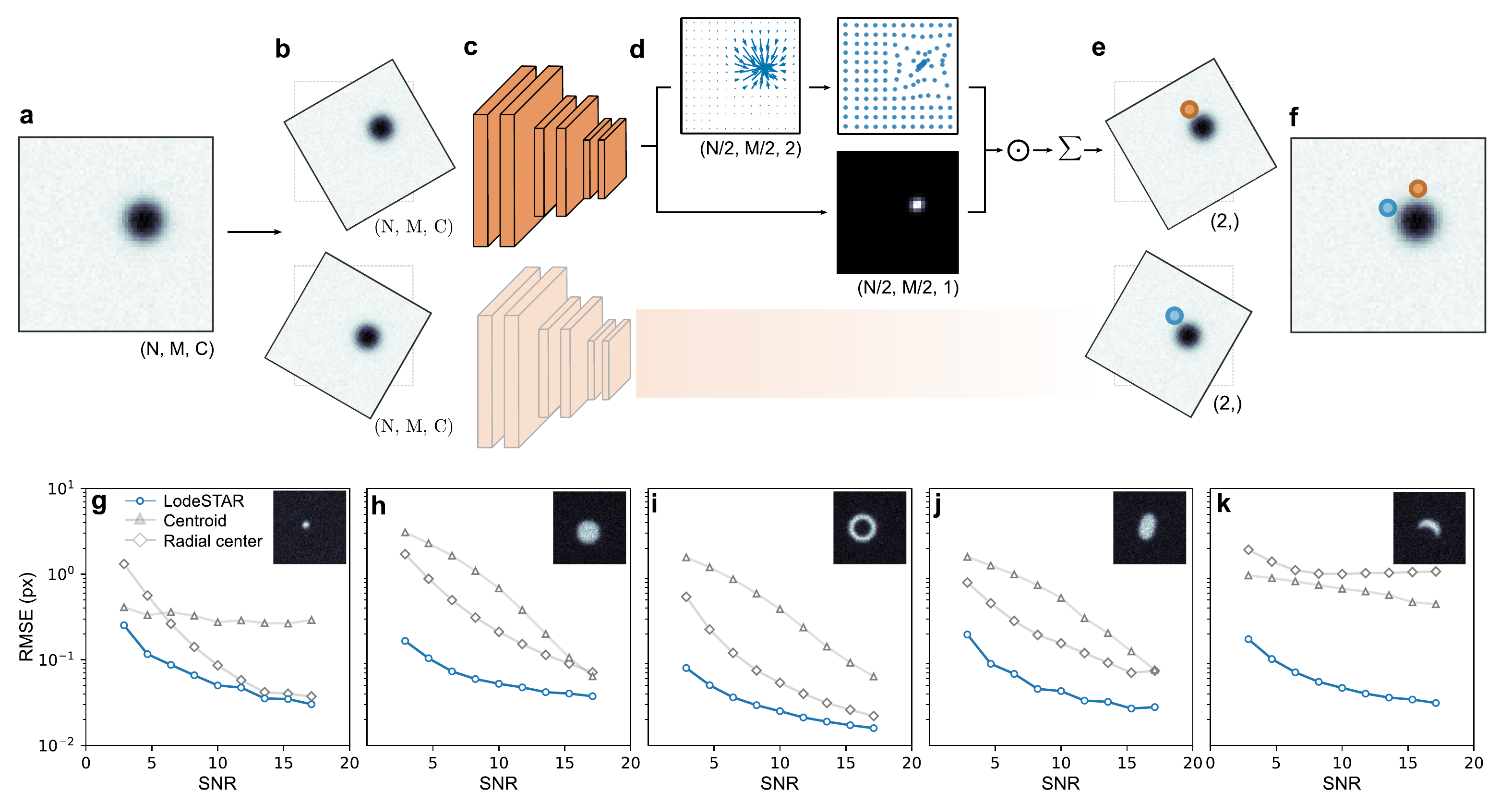}
    \caption{
    \footnotesize
    \textbf{LodeSTAR single-shot training and performance.} 
    \textbf{a} Example image of a single particle used to train the neural network ($N \times M$ pixels, $C$ color channels). 
    \textbf{b} Two copies of the original image transformed by translations and rotations. 
    \textbf{c} The transformed images are fed to a convolutional neural network.
    \textbf{d} The neural network outputs two tensors (feature maps), each with $N/2 \times M/2$ pixels: 
    One (top) is a vector field where each pixel represents the distance from the pixel itself to the particle (top left, blue arrows), which is then transformed so that each pixel represents the distance of the particle from the center of the image (top right, blue markers).
    The other tensor (bottom) is a weight map (normalized to sum to one) corresponding to the contribution of each element in the top feature map to the final prediction.
    \textbf{e} These two tensors are multiplied together and summed to obtain a single prediction of the position of the particle for each transformed image. 
    \textbf{f} The predicted positions are then converted back to the original image by applying the inverse translations and rotations. The neural network is trained to minimize the distance between these predictions. 
    \textbf{g-k} LodeSTAR performance on $64\,\rm{px}\times 64\,\rm{px}$ image containing different simulated particle shapes: \textbf{g} point particle, \textbf{h} sphere, \textbf{i} annulus, \textbf{j} ellipse, and \textbf{k} crescent. 
    Even though LodeSTAR is trained on a single image for each case (found in the corresponding inset), its root mean square error (RMSE, blue circles) shows that it outperforms traditional methods based on the centroid \cite{Crocker1996} (gray triangles) or radial symmetry \cite{Parthasarathy2012} (gray diamonds), especially at low signal-to-noise ratios (SNRs). 
    Interestingly, even in the crescent case \textbf{k}, where there is no well-defined particle center, LodeSTAR is able track it to within a tenth of pixel.}
    \label{fig1}
\end{figure}

LodeSTAR is designed to exploit the symmetries inherent in particle tracking. 
For example, even if we do not know the absolute position of a particle, we can say for sure that, if we translate the particle image by a certain amount, its position gets also translated by the same amount --- and similarly for rotations and reflections. This is known as an equivariance.
In fact, for a particle whose image has a well-defined center of symmetry, such as that shown in Figure~\ref{fig1}a, a prediction of the position of the particle that is equivariant to the Euclidean group (translations, rotations, and reflections) can be shown to necessarily locate the center of the particle (see Methods, `Theory of Geometric Self-Distillation'').

LodeSTAR trains a neural network to achieve equivariance, i.e., an exact correspondence between the transformation applied to its input image (e.g., a translation by a certain amount) and the effect this has on the output prediction (e.g., a translation of the predicted particle position by the same amount).
We start with an image of a single particle (Figure~\ref{fig1}a). 
By applying a transformation from the Euclidean group to this image, we create several transformed images (e.g., the two images in Figure~\ref{fig1}b). 
A neural network (Figure~\ref{fig1}c) predicts the position of the particle within each image (Figure~\ref{fig1}d, which will be more thoroughly explained in the next paragraph), yielding two predictions of the $x$ and $y$ coordinates of the particle, one for each transformation (Figure~\ref{fig1}e). 
To compare these predictions, we back-transform the predicted coordinates to determine their corresponding positions in the original image (Figure~\ref{fig1}f). 
Then, if the neural network is equivariant to the applied transformations, the two predictions should perfectly overlap, while any deviation between these two prediction can be used to train the neural network (i.e., to adjust the neural network by updating internal weights to minimize the distance between the two predictions). 

We now turn our attention to the details of the neural network employed by LodeSTAR.
While most architectures are compatible with LodeSTAR, it is convenient to use a completely translation-equivariant neural network (e.g., a fully convolutional neural network).
This presents a two-fold advantage.
First, having inherent translation-equivariance, the neural network does not need to learn the equivariance during training, which significantly reduces the required complexity of the model (while rotation-equivariance and reflection-equivariance still need to be learned).
Second, as will become clear in the following sections, this will permit us to use the network also to track multiple particles while still training on a single particle image.

As schematically shown in Figure~\ref{fig1}c (details are described in the Methods, ``Neural network architecture''), we use a fully convolutional neural network, with two convolutional layers, one max-pooling layer, and seven additional convolutional layers, the last of which outputs three channels, $\Delta x$, $\Delta y$, and $\rho$.
The outputs of this neural network are further analyzed, as shown Figure~\ref{fig1}d (this analysis also needs to be translation-equivariant).
The vectors $(\Delta x, \Delta y)$ estimate the distance from each pixel to the particle (blue arrows in upper left panel of Figure~\ref{fig1}d; far away from the particle, the length of these vectors goes to zero because of the limited receptive field of the convolutional neural network, indicating that it does not see a particle in its input data). By adding to each of these vectors the respective pixel position, we retrieve a map of predictions of the particle position relative to the upper left corner of the image (blue markers in the upper right panel of Figure~\ref{fig1}d). 
The $\rho$ channel provides a weight map
(normalized to sum to one) corresponding to the probability of finding the center of
the particle near each pixel (bottom panel in Figure~\ref{fig1}d).
The final prediction of the particle position is obtained by an average of the estimated particle positions weighted by the weight map.

LodeSTAR manages to train this network using a single image of the particle. Moreover, thanks to the small size of the neural network, it can be trained fully from scratch in an order of $10^4$ mini-batches, which takes a few minutes, even without a graphics processing unit. Finally, LodeSTAR can be trained with small batch-sizes, and as such needs less than a gigabyte of runtime memory (see details in the Methods, ``Neural network training'').

\subsection*{Tracking a single particle}

We start by considering the performance of LodeSTAR on the simplest case: a point particle (e.g., the image of a single molecule obtained from a fluorescence microscope). 
We simulate $10^4$ images of point particles with signal-to-noise ratio (SNR) between $2$ and $20$ using the Python library DeepTrack 2.1 \cite{Midtvedt2021Quantitative}.
We use a single one of these images (inset in Figure~\ref{fig1}g, ${\rm SNR}=10$) to train LodeSTAR, using 5000 mini-batches of 8 samples.
LodeSTAR achieves a sub-pixel root mean square error (RMSE) for all SNRs and a ${\rm RMSE}<0.1\,{\rm px}$ for ${\rm SNR}>5$ (blue circles in Figure~\ref{fig1}g).
Strikingly, LodeSTAR performs well also far from the SNR at which it is trained. 

In order to assess the performance of LodeSTAR, we compare it with two of the most standard particle tracking methods, i.e., the centroid method \cite{Crocker1996} (gray triangles in Figure~\ref{fig1}g) and the radial-center method \cite{Parthasarathy2012} (gray diamonds in Figure~\ref{fig1}g). 
While the radial center method achieves sub-pixel accuracy over the whole range of SNRs, it is consistently outperformed by LodeSTAR.
The radial-center method approaches LodeSTAR's performance for high SNRs, but is not competitive at lower SNRs, even underperforming the centroid method.
Overall, these results are consistent with the literature, where deep-learning-based methods have been shown to outperform traditional methods especially at low SNRs \cite{Midtvedt2021Quantitative,Newby2018,Helgadottir2019}.

In contrast to traditional methods that are optimized for certain sets of particle shapes, LodeSTAR can track particles of arbitrary shapes.
We show some examples in Figures~\ref{fig1}h-k, where we present the results for a sphere (Figure~\ref{fig1}h), an annulus (Figure~\ref{fig1}i), an ellipse (Figure~\ref{fig1}j), and a crescent (Figure~\ref{fig1}k).
Overall, LodeSTAR achieves a sub $0.1\,\rm{px}$ error for the majority of the SNR-range in all cases, outperforming the standard methods by up to an order of magnitude.
While we can expect LodeSTAR to perform well on symmetric objects, such as the sphere (Figure~\ref{fig1}h) and the annulus (Figure~\ref{fig1}i), it is an important confirmation of the generality of LodeSTAR that it also works for non-symmetric particles, such as the ellipse (Figure~\ref{fig1}j), which only has two axes of symmetry instead of full radial symmetry, and the crescent (Figure~\ref{fig1}k), which has only one axis of symmetry.
The latter case is particularly interesting, because the standard methods perform rather poorly (${\rm RMSE}>1\,{\rm px}$), while LodeSTAR retains a sub $0.1\,\rm{px}$ error, on-par with the symmetric cases (since the center of a crescent moon is not uniquely defined, we accept any position along the axis of symmetry to be a valid prediction as long as it is consistent across all images). Video~1 demonstrates LodeSTAR tracking the particles at various orientations and noise levels.

We turn now our attention to the discriminative power of LodeSTAR, i.e., its ability to learn about the specific shape used in its training.
Such discriminative power is important for heterogeneous samples or just to minimize false positives.
In order to do this, we consider the distribution of predicted positions of the particle in view preceding the pooling operation, i.e., the feature map (Figure~\ref{fig1}d). Specifically, we measure the self-consistency of the model by calculating the weighted variance of the feature map. We expect that the neural network is highly self-consistent when evaluated on images similar to the training data, while less so when evaluated on something distinctly different.
As can be seen in Table~\ref{tab1}, the variance is several orders of magnitude lower when the model analyzes an image drawn from the distribution on which it has been trained.
This demonstrates that LodeSTAR acts differently when presented data similar to the training distribution compared to data from another distribution, which clearly shows that LodeSTAR has acquired discriminative power. 

\begin{table}[]
    \centering
    \begin{tabular}{r|l|l|l|l|l}
        \diagbox{Model}{Particle} & Point & Sphere & Annulus & Ellipse & Crescent moon\\
        \hline
        Point & \textbf{0.02} & 31.44 & 70.47 & 12.82 & 1590.37 \\
        Sphere & 857.96 & \textbf{0.05} & 58.66 & 1.13 & 1025.23 \\
        Annulus & 58.60& 26.94 & \textbf{0.07} & 27.45 & 750.94 \\
        Ellipse & 758.71 & 0.85 & 75.16 & \textbf{0.12} & 21.77 \\
        Crescent & 13.97 & 11.47 & 18.32 & 10.10 & \textbf{0.18} \\
    \end{tabular}
    \caption{
    \textbf{LodeSTAR displays discriminative power.}
    Each model trained by LodeSTAR is significantly more self-consistent (lower average weighted variance) when presented data from its training distribution (bold font), compared to when it is evaluated on different data. This demonstrates that LodeSTAR has acquired some discriminative power.
    }
    \label{tab1}
\end{table}

\subsection*{Tracking multiple particles}

\begin{figure}[h!]
    \centering
    \includegraphics[width=\columnwidth]{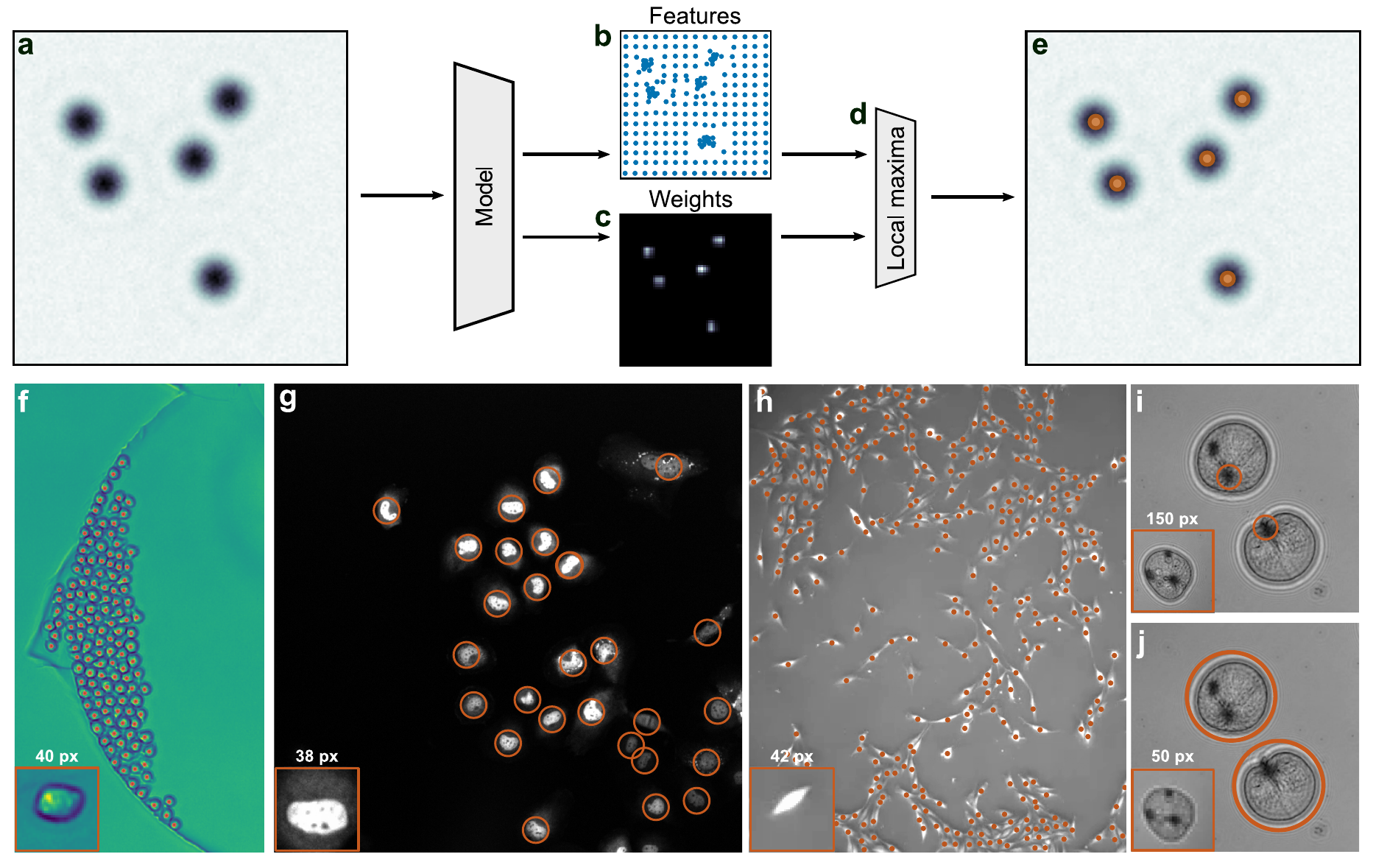}
    \caption{
    \footnotesize
    \textbf{LodeSTAR analysis of images with multiple particles.} 
    \textbf{a} Example image with multiple particles to be analyzed by LodeSTAR (LodeSTAR is still trained on a single particle image, as described in Figure~\ref{fig1}). 
    \textbf{b} LodeSTAR returns clustered predictions of particle positions and \textbf{c} a weight map representing the likelihood  of finding a particle near each pixel. 
    \textbf{d} An estimation of the local density of particle detections is multiplied by the weight map to obtain a detection map,  whose local maxima are considered particle detections (orange markers in \textbf{e}).
    \textbf{f}-\textbf{i} Examples of applications of LodeSTAR to experimental data that present different challenges. In all cases, LodeSTAR is trained on the single crop shown in the respective inset and then applied to the whole time-series. See also the corresponding Videos~2-6.
    \textbf{f} LodeSTAR finds the positions of mouse hematopoietic stem cells (red markers), achieving a $99.3\%$ true positive rate and a $0.9\%$ false discovery rate despite the dense sample (data from \cite{Ulman2017}). 
    \textbf{g} LodeSTAR identifies human hepatocarcinoma-derived cells (red circles), achieving a $92.8\%$ true positive rate and a $2.8\%$ false discovery rate, despite the high variability between cells (data from \cite{Ulman2017}). 
    \textbf{h} LodeSTAR tracks pancreatic stem cells (red markers), achieving a  $92.4\%$ true positive rate and a $8\%$ false discovery rate, despite the densely packed sample and the high variability between cells (data from \cite{Ulman2017}). 
    \textbf{i} LodeSTAR tracks the plankton \emph{Noctiluca scintillans}. In this case, LodeSTAR tracks the optically dense area of the tentacle attachment point (red circles). 
    \textbf{j} Interestingly, if the data is downsampled by a factor of 3 (so that the training image is $50\,{\rm px} \times 50\,{\rm px}$ instead of $150\,{\rm px} \times 150\,{\rm px}$) before training and evaluation, the model finds the cell as a whole. 
    }
    \label{fig2}
\end{figure}

Thanks to the translation-equivariant design of the neural network, LodeSTAR trained on a single particle image as described in the previous section can immediately be used to track multiple particles without any additional training.
In fact, since the receptive field of the convolutional network is limited, additional particles in view (such as in Figure~\ref{fig2}a) are analyzed largely independently by the neural network. 
Taking advantage of these observations, we can bestow LodeSTAR with the capability to detect multiple particles, simply by removing the weighted global pooling layer (i.e., the final multiplication in Figure~\ref{fig1}e) and operating directly on the feature-maps themselves, i.e., on the predicted particle position map (Figure~\ref{fig2}b) and on the weight map (Figure~\ref{fig2}c).
By multiplying a measure of the local density of particle positions by the weight map (Figure~\ref{fig2}d), we obtain a detection map whose local maxima represent the positions of the detected particles.
After detection, the exact location of the particle is determined by a weighted average of the local region around the detection, analogous to the single-particle case, finally yielding a successful tracking of the image (Figure~\ref{fig2}e).
See also details in Methods, ``Particle detection criteria''.

\subsection*{Validation with experimental data}

We now apply LodeSTAR to various experimental images with multiple particles, which present different challenges (Figures~\ref{fig2}f-j).
We highlight that in all cases we train LodeSTAR on a single image (shown in the insets in the lower left corners in Figures~\ref{fig2}f-j) and then apply it to the whole image. 
In all cases, the objects are dividing over time, resulting in a large range of densities and morphologies that LodeSTAR needs to handle. Videos visualizing the tracked images can be found for each case in the supplementary material (Videos~2-6).

First (Figure~\ref{fig2}f), we consider a dense sample. 
We analyze an experimental time-series of images of densely-packed mouse stem-cells.
Even though the cells are very densely packed (in fact, they often touch each other), LodeSTAR manages to accurately identify their positions (red markers in Figure~\ref{fig2}f), achieving a true positive rate of $99.4\%$ and a false discovery rate of $0.9\%$. 
Strikingly, the ability to separate closely packed cells is completely emergent, since the neural network is never shown images with multiple cells during its training.

Then (Figure~\ref{fig2}g), we consider an experimental scenario where the cells vary highly both in morphology and in intensity.
We track the nuclei of human hepatocarcinoma-derived cells, showing the identified cell positions as red circles.
Despite being a highly dynamic dataset with a very large variability between cells, LodeSTAR achieves a true positive rate of $92.8\%$ and a false discovery rate of $2\%$. 
Furthermore, it should be noted that most missed detections stem from not immediately detecting cell division events present in the data. 

In Figure~\ref{fig2}h, we consider the doubly-challenging case of densely packed cells that vary in morphology and intensity. We analyze a time-series of pancreatic stem cells, showing the identified cell positions as red dots.
Despite the cells being highly packed and varying significantly over instance and time, LodeSTAR achieves a $92.4\%$ true positive rate and a $8\%$ false discovery rate. However, it should be noted that a large portion of the false discoveries originate from errors in the ground truth provided by the Cell Tracking Challenge\cite{Ulman2017}.

Finally, we consider some more complex objects with significant internal structure, some \emph{Noctiluca scintillans}, large ($400\,{\rm \upmu m}$ to $1500\,{\rm \upmu m}$) single-celled dinoflagellate planktons (Figures~\ref{fig2}i-j, see also Methods, ``Plankton preparation and imaging''). 
LodeSTAR works by identifying some characteristic feature of the object to be tracked. If this object is symmetric, the tracked feature is its center. 
Since the planktons have a complex internal structure, LodeSTAR does not necessarily find their geometrical center, but it nonetheless consistently identifies some specific feature of their internal structure.
Furthermore, we can expect this feature to depend on the details of the neural network and its training.
In fact, when LodeSTAR is trained on the inset of Figure~\ref{fig2}i ($150\,{\rm px}\times 150\,{\rm px}$), it learns to consistently identify the region where the tentacle attach and organelles aggregate.
In contrast, when we downsample the images by a factor of 4 so that the training is made on the $50\,{\rm px}\times 50\,{\rm px}$ figure shown in the inset of Figure~\ref{fig2}j, LodeSTAR consistently finds the plankton as a whole.
In this way, it is possible to tune the scale of the tracking performed by LodeSTAR.

\subsection*{Exploiting additional symmetries}

LodeSTAR can exploit additional symmetries to extend the range of particle properties that it can measure. 
Off-axis holography is a prime example of an imaging modality ripe with additional symmetries \cite{Tahara2018}. Unlike ordinary brightfield microscopy where only the intensity of the incoming wave is imaged, off-axis holography permits access to the entire complex electromagnetic field. Consequently, it is possible to infer quantitative information about the imaged objects by measuring and manipulating the Fourier representation of the image. 
As we will see in the following sections, when analyzing holographic microscopy images, we can exploit the Fourier propagation symmetry to detect the axial position and the signal strength scaling symmetry to determine the particle polarizability.
Implementation-wise, additional symmetries are encoded as additional channels of the intermediate feature-map, and are trained using their own set of equivariances in the same manner as for the in-plane position we have discussed in detail until now. 

\subsection*{3D tracking exploiting Fourier propagation symmetry}

\begin{figure}[b!]
    \centering
    \includegraphics[width=\columnwidth]{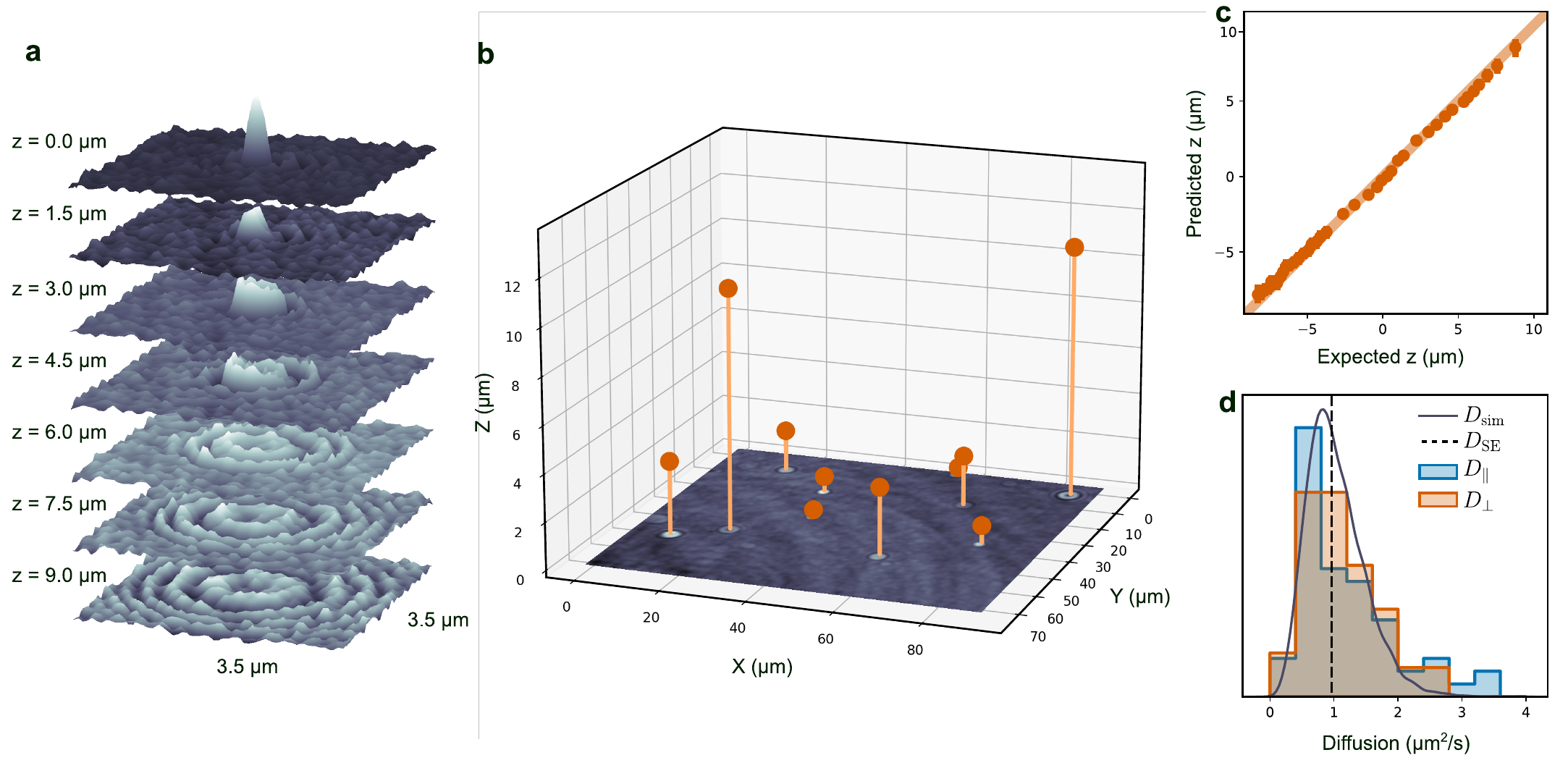}
    \caption{
    \footnotesize
    \textbf{LodeSTAR measurement of 3D positions exploiting Fourier propagation symmetry.} 
    \textbf{a} Imaginary part of the hologram of a $228\,{\rm nm}$ radius polystyrene particle numerically Fourier-propagated to different axial distances from the focal plane.
    \textbf{b} LodeSTAR exploits this Fourier propagation symmetry to learn how to track particles in three dimensions (orange positions).
    \textbf{c} The vertical position estimated by LodeSTAR agrees well with the expected position acquired using a traditional approach described in \cite{Midtvedt2021}. See also Video~7.
    \textbf{d} The distributions of the in-plane $xy$-diffusion (blue histogram) and the axial $z$-diffusion (orange histogram) of the particles show a strong peak at at the expected diffusion (dashed black line), and agree well with the expected theoretical distribution obtained by calculating the diffusion constant of $10^4$ synthetic traces (solid black line).}
    \label{fig3}
\end{figure}

A natural extension of two-dimensional tracking is to incorporate the third axial dimension, normal to the imaging plane.
This can give crucial insight into the full volume dispersion of objects, as well as provide more data to calculate statistical measures about the objects' motility, such as their diffusion.
An example of (the imaginary part) of an holographic image of a particle (a 228-nm-radius polystyrene sphere) is shown by the top slice ($z = 0\,{\rm \upmu m}$) in Figure~\ref{fig3}a.
The holographic image can be propagated to different planes (i.e., different axial positions from the focal plane) by employing Fourier transforms, as shown by the other slices in Figure~\ref{fig3}a.
This provides an equivariance that LodeSTAR can learn, similar to the equivariances in the plane.
By training on the image in the top slice of Figure~\ref{fig3}a, LodeSTAR learns to track the polystyrene spheres in 3D space, as shown in Figure~\ref{fig3}b, where the measured vertical position is visualized as the distance above the image.  Note that the refractive indices of the medium ($n_{\rm medium}$) and the immersion oil ($n_{\rm oil}$) change the apparent axial position of the particle\cite{sibarita2005deconvolution}. We correct this by multiplying the measured axial position by a factor $n_{\rm oil}/n_{\rm medium} = 1.128$\cite{sibarita2005deconvolution}.

As a verification, we compare the predicted vertical position with a traditional focusing approach where the vertical position is found by iteratively refocusing a region near a detection until a focusing criterion is met \cite{Midtvedt2021}. 
As shown in Figure~\ref{fig3}c, the two methods are in very good agreement. Moreover, we calculate the standard deviation of the error of the two methods based on the covariance estimate proposed in Ref.~\cite{Vestergaard2014diffusion}: LodeSTAR outperforms the classical approach significantly, reaching $\sigma_z=105\,{\rm nm}$ compared to $\sigma_z=231\,{\rm nm}$.

As a second verification, independent of the traditional approach, we compare the predicted diffusion constants of each particle trace, calculated either from its in-plane or axial movement. 
Since the diffusion constant is a statistical measure of the thermal motion of the particle, the diffusion of individual particles in different directions may not agree, while the ensemble of all particles should. As such, we compare the distribution of measured diffusion constants in Figure~\ref{fig3}d for in-plane movement and axial motion. 
We find that both in-plane diffusion and axial diffusion are distributed similarly, and that both show a peak at the expected diffusion of $D_{\rm SE} = 0.97\,{\rm \upmu m^2 s^{-1}}$. Moreover, we find that the distributions closely agree with the expected distribution of diffusion constants obtained by simulating $10^4$ traces. 

\subsection*{Particle polarizability exploiting signal strength symmetry}

\begin{figure}[h!]
    \centering
    \includegraphics[width=\columnwidth]{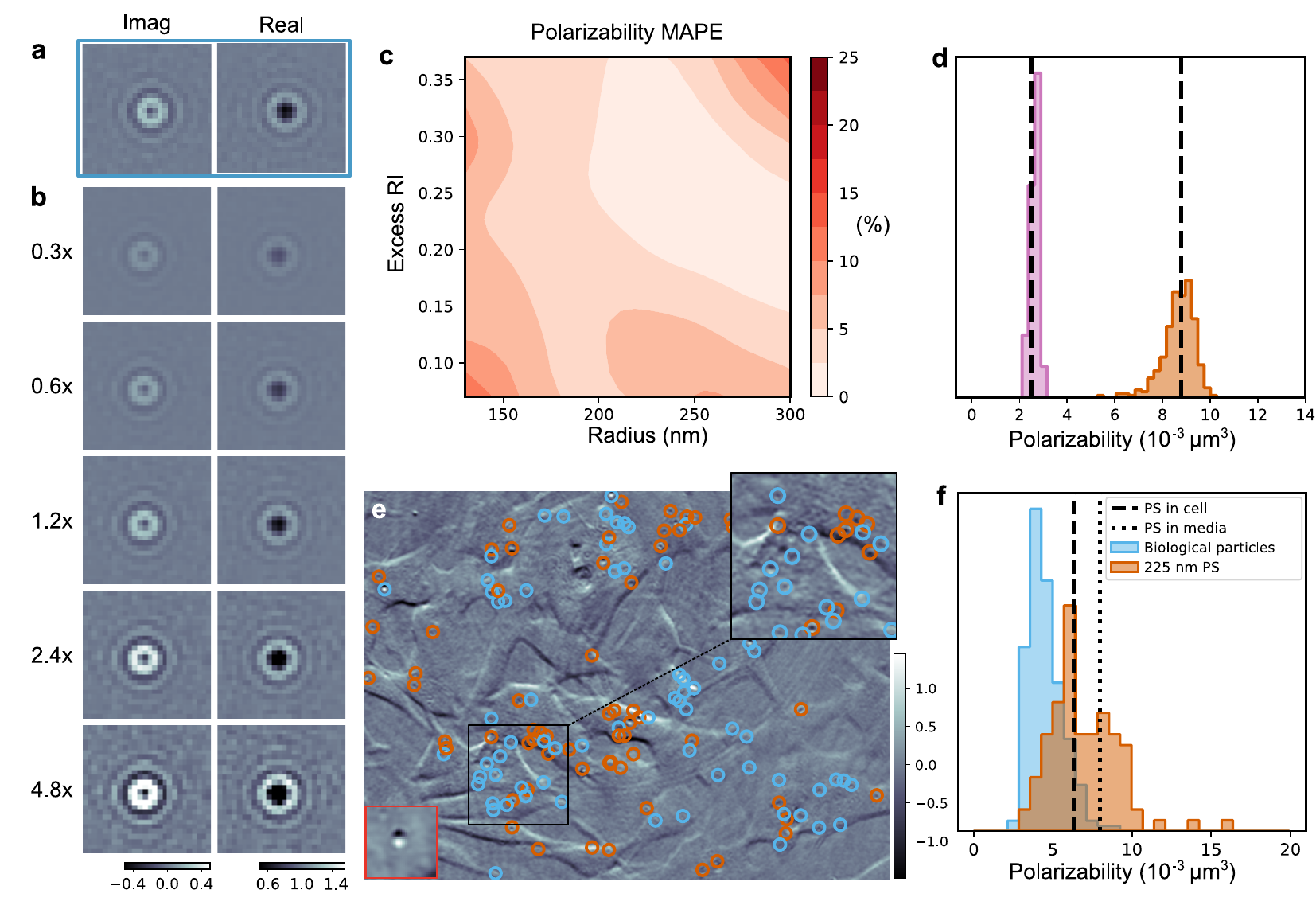}
    \caption{
    \footnotesize
    \textbf{LodeSTAR measurement of particle polarizability exploiting signal strength symmetry.} 
    \textbf{a} Real and imaginary part of a simulated holographic image of a sphere (radius $228\,{\rm nm}$, refractive index 1.58), \textbf{b} their versions with numerically rescaled signal strengths. 
    \textbf{c} Despite being trained on a single particle (radius $228\,{\rm nm}$, refractive index 1.58), the mean absolute percentage error (MAPE) of the predicted polarizability remains below $10\%$ for a wide range of particle sizes and refractive indices.
    \textbf{d} In an experimental bi-dispersed sample, LodeSTAR accurately estimates the polarizability of the $150\,{\rm nm}$ population of polystyrene particles, even though it is trained on an image from the $228\,{\rm nm}$ population. 
    \textbf{e} The imaginary part of an example position-modulated holography image of fluorescent polystyrene particles (radius $225\,{\rm nm}$) suspended inside and around SH-SY5Y human neuroblastoma cells imaged through an off-axis holography microscope. LodeSTAR, trained on a single image (bottom left), learns to detect and measure the fluorescent particles (orange markers) as well as the non-fluorescent intracellular particles (blue markers). See also Video~8.
    \textbf{f} The distributions of the measured polarizability of the particles and the biological particles are drawn from two distinct distributions, indicating that we successfully separate the added polystyrene particles from the biological particles. The peak of the distribution matches the expected polarizability of polystyrene inside of cells (dashed line), a less prominent peak near the expected polarizability of polystyrene outside of cells (dotted line).}
    \label{fig4}
\end{figure}

The holographic image of a particle carries information not only about the particle position, but also about its morphology and composition. 
For example, the real part of the polarizability of an object (which henceforth will be referred to as just the polarizability for convenience) is proportional to the integrated phase acquired by the light passing through the particle, which is particularly relevant for biological objects, where refractive index and density are strongly correlated. In other words, for biological materials, it is possible to directly translate the polarizability of a particle to its mass \cite{Zangle2014}. The integrated phase, in turn, increases roughly proportional to the amplitude of the scattered light for small non-absorbing particles.
Thus, there is an equivariance between the polarizability and the scale of the signal that LodeSTAR can learn. 

Figure~\ref{fig4}a shows the real and imaginary parts of a simulated holographic image of a particle with radius $228\,{\rm nm}$ and refractive index $1.5$. 
By multiplying these images by a factor (Figure~\ref{fig4}b), we can alter the signal scale and, therefore, the polarizability of the imaged particle. 
LodeSTAR is trained to estimate the logarithmic difference between the scale factors, in addition to the particle position.
We remark that this equivariance does not constrain the absolute scale of the polarizability, which thus needs to be calibrated against some observation of known polarizability.

In Figure~\ref{fig4}c, we evaluate the trained LodeSTAR on simulated particles, varying their radius and their refractive index. 
The mean absolute percentage error remains below $10\%$ for the majority of the considered range, only increasing for very low signal observations where noise would corrupt most of the signal. 
This is comparable to the findings of other methods, and yields accurate determinations by averaging over several observations of the same particle \cite{Altman2020,Midtvedt2021}. 

In Figure~\ref{fig4}d, we validate LodeSTAR's ability to measure the polarizability of particles in experimental data.
We consider a bi-dispersed sample of $150\,{\rm nm}$ radius and $228\,{\rm nm}$ radius polystyrene particles imaged through a holographic microscope. 
LodeSTAR is trained on one observation from the $228\,{\rm nm}$ population and is subsequently used to predict the polarizability of all particles in the sample. By calibrating against the $228\,{\rm nm}$ population, LodeSTAR successfully identifies the $150\,{\rm nm}$ population with high accuracy.

Having verified that LodeSTAR can reliably determine the polarizability of particles, we consider a biological sample with $225\,{\rm nm}$ radius green-fluorescent polystyrene beads that were incubated with human neuroblastoma cell from the SH-SY5Y cells-line (Figure~\ref{fig4}e, see Methods, ``Human neuroblastoma cell sample preparation''). The particles are simultaneously imaged with sample-position-modulated holography and fluorescence (see Methods, ``Holographic imaging''). This allows us to classify detections as polystyrene particles (orange markers, which are fluorescent) or biological aggregates (blue markers, which are not fluorescent). As can be seen in the zoomed-in region in Figure~\ref{fig4}e, the signal is extremely low compared to the background. As such, one can expect spurious detections. We filter these out by only considering observations that could be linked over time for at least $40$ frames, disregarding observations that arose from random noise. We find that the majority of detections are co-located with the cells. Further, studying the time-series (Video~8) the particles move in unison with the cells, supporting the premise that some of the particles have been taken up by the cells. See also details in Methods, ``Measuring particle polarizability''. 

Finally, we study the distribution of polarizability for the particles and the biological matter in Figure~\ref{fig4}e. We find a much broader distribution for the polystyrene particles than the bi-dispersed case, which is expected if some of the particles are measured inside of the cells, since the intracellular medium has a higher refractive index than the surrounding medium (1.38 compared to 1.33 \cite{YangCyto}) and consequently yields slightly lower polarizability of the particle. Moreover, particles inside the cells are likely to be coated with biological material, further broadening the distribution.
The peak of the distribution aligns well with the expected polarizability of PS inside the cell, given a cytoplasmic refractive index of 1.38\cite{YangCyto}. However, a second peak near the polarizability of PS in water suggests that a significant portion of the particles have not entered the cells, which agrees with other cell--particle uptake experiments \cite{Ashraf2020}.

The distribution of polarizability of biological particles is significantly narrower and peaks at just under $0.01\,{\rm \upmu m^3}$. It should be noted that the detection of objects with polarizability lower than $0.006\,{\rm \upmu m^3}$ is not reliable due to the low signal, which means that the lower end of the distribution should be analyzed with some caution. Regardless, the two distributions are clearly distinct, indicating that they represent two separate physical properties of the sample.

\section*{Discussion and conclusions}

We have developed a method, named LodeSTAR, that exploits the inherent symmetries of a problem to enable label-free training of neural networks using tiny datasets. We have demonstrated this capability by training neural networks to track objects in a broad range of simulated and experimental scenarios. Moreover, we have shown that LodeSTAR can quantitatively measure the objects in terms of their axial position and their polarizability. The software together with the source code and all the examples in this work are made publicly available through the DeepTrack 2.1 GitHub repository \cite{deeptrackgithub}.

Compared to traditional approaches, we are able to achieve a better tracking performance in terms of sub-pixel accuracy, while generalizing to more arbitrary morphologies. 
Moreover, unlike established methods for object tracking using deep learning, which commonly require hundreds if not thousands of annotated images for training \cite{Midtvedt2021Quantitative}, LodeSTAR successfully tracks difficult experimental data using just a single datapoint for training. 
On top of this, LodeSTAR can be trained on an ordinary computer with no hardware acceleration (e.g., graphics processing units) within a few minutes.

As a novel approach to an established problem, we see several future lines of inquiry. Foremost, we expect the existence of many additional symmetries not considered in this work, particularly symmetries in the experimental design (such as the symmetries that allowed us the measure the axial position and polarizability of nanoparticles). We also expect that incorporating techniques from other areas of deep learning to be fruitful; for example, using active learning to optimize performance from low amounts of human supervision. Regarding the development of the technique itself, we consider the development of techniques to further improve the specificity of the model, such as leveraging negative labels to indicate what \emph{not} to track. Finally, LodeSTAR may be used for segmentation, similarly to the results of achieved with the DINO (self-distillation with no labels) algorithm \cite{caron2021emerging}, leveraging the fact that only regions within the object are informative to its position.

The comparably low barrier of entry permits rapid creation of a custom tracking solution, requiring little-to-no expertise from the user. Additionally, by side-stepping the need for synthetic data, LodeSTAR opens up the possibility to train high-quality models to analyze data that is difficult to reproduce synthetically, without relying on fallible human annotation.

\section*{Methods}

\subsection*{Theory of geometric self-distillation}

To ensure that the training converges to the physically correct solution (e.g., identifying the centroid of the particle), the symmetry group used to train LodeSTAR needs to have a set of properties. We will describe them here.

Let $X$ be the set of possible input images, $Y$ be the set of possible outputs, $f$ and $h$ be functions $h:X \to Y$, and $G$ be a group of transformations acting on both $X$ and $Y$. 
(In particle tracking, $X$ is all possible images of the particle, $Y$ is all possible particle positions $\mathds{R}^2$, $f$ is the ground-truth function that returns the particle centroid, $h$ is the neural network that need to be trained to return the particle centriod, and $G$ is the Euclidean group consisting of translations, rotations, and reflections.) 

We can show that $f$ is necessarily identical to $h$ for $\forall x\in X$, if the following set of assumptions hold:
(1) $f(gx) = gf(x)$ and $h(gx) = gh(x)$ $\forall g \in G, x\in X$ --- i.e., both $f$ and $h$ are equivariant to $G$. 
(2) For any $x_1$ and $x_2$ in in $X$, there exists a $g\in G$ such that $g x_1 = x_2$ (known as transitivity in group action theory). 
(3) There exists two distinct $g_1, g_2 \in G$ such that $g_1x = g_2x$ for some $x\in X$, i.e., $G$ is not free. 
(4) There exist an $x \in X$ and a g $\in G$ that fixes $x$ for which $f(x)$ is the only fixed point. 
(In particle tracking: 
(1) asserts that the centroid of the particle ($f$) and the neural network ($h$) are equivariant to the Euclidean group. $f$ is true by definition, $h$ is true by training. 
(2) asserts that the image of a particle in one position and orientation should be transformable to any other position and orientation using a combination of Euclidean transformations. 
(3) asserts that there exists some non-trivial transformation that leaves the image of the particle unchanged. For example, rotating a square 90 degrees does not change it. 
(4) asserts that there exists a transformation $g$, such that $gx = x$ and $gy = y$ hold if and only if $y$ is the centroid of the particle whose image is $x$.)

Since per (1) $h$ is equivariant, $h(gx) = gh(x)$. Further, due to (3), $gx = x$, and consequently $h(gx) = h(x)$. Together, we find that $gh(x) = h(x)$.  Finally, due to (4), there exists a $g^\prime$ and $x^\prime$ such that $g^\prime y = y$ is true only if $y = f(x^\prime)$. Consequently, $g^\prime h(x) = h(x)$ can only be true if $h(x^\prime) = f(x^\prime)$. Now, consider some arbitrary $g \in G$. We find that $f(gx^\prime) = gf(x^\prime) = gh(x^\prime) = h(gx^\prime)$, where the first and third equality signs are given by equivariance. As such, if $f(x^\prime) = h(x^\prime)$ for some $x^\prime$, then $f(x) = h(x)$ for all $x$ reachable by acting $g$ on $x$. Due to (2), all $x \in X$ are reachable from $x^\prime$ through actions from $G$. Consequently, $f(x) = h(x)$ for all $x\in X$.  

Assumptions (1) and (2) ensure that the network will consistently track the particle, while (3) and (4) ensure that exactly the centroid is tracked. As such, even if (3) does not hold (as is the case for asymmetrical particles), the particle will still be tracked consistently.

\subsection*{Neural network architecture}

The neural network consists of three $3 \times 3 \times 32$ convolutional layers with ReLU activation, followed by a $2 \times 2$ max-pooling layer, followed by eight $3 \times 3 \times 32$ convolutional layers with ReLU activation, and finally by a single $1 \times 1 \times 3$ convolutional layer with no activation. 
Splitting the output channels into $\Delta x$, $\Delta y$ and $\rho$ respectively, LodeSTAR first calculates
\[
x_{i,j} = \Delta x + ik - \frac{N}{2k},
\]
\[
y_{i,j} = \Delta y + jk - \frac{M}{2k},
\]
where $N$ is the size of the input along the first dimension, and $M$ is the size of the tensor along the second dimension, and $k$ is a scale factor relating the size of the input to the output ($k=2$ for the proposed architecture). 
During runtime, the weights are normalized as
\[
w_{i, j} = \mathrm{S}(\rho_{i, j}),
\]
where $\mathrm{S}(\cdot)$ is the sigmoid function used to constrain individual elements between 0 and 1, which makes choosing a detection threshold for multi-particle tracking easier.

\subsection*{Neural network training}

During training, the weights are normalized differently, namely:
\[
w_{i, j} = \frac{\mathrm{D}[\mathrm{S}(\rho_{i, j}), 0.01]  + \epsilon}{\epsilon MN + \sum \mathrm{D}[\mathrm{S}(\rho_{m, n}), 0.01]},
\]
where $\epsilon$ is some small value ($10^{-6}$), $\mathrm{D}[\cdot, 0.01]$ is a dropout with a dropout-rate of $1\,\%$. 
The dropout avoids the solutions where a single element is large and the rest are small, increasing the robustness of the network. The epsilon adds non-linearity during training, forcing the network to utilize the entire span of the sigmoid function.
Further, the network strives to minimize the weight of non-informative regions. The only way to achieve this is to increase the normalization factor $ \sum W^\prime_{i, j}$, which is maximized by predicting large values at informative regions. During evaluation, the weight is not normalized, in order to make it independent of the number of objects in view.

We employ two loss functions during training. The first is a loss between the untransformed predictions and their mean. The second is an internal consistency loss, which, for the case of the particle centriod prediction, is calculated as
\[
\mathcal{L}^{x}_{i,j} = \left\lvert x_{i,j} - \sum x w\right \rvert w_{i,j},
\]
\[
\mathcal{L}^{y}_{i,j} = \left\lvert y_{i,j} - \sum y w\right \rvert y_{i,j},
\]
\[
\mathcal{L_{C}} = \sum \mathcal{L}_{i,j}^{x} + \mathcal{L}_{i,j}^{y}.
\]
In the case where additional features are predicted (e.g., $z$-position and polarizability), the corresponding loss term is calculated in the same way as $\mathcal{L}^{x}_{i,j}$ and $\mathcal{L}^{y}_{i,j}$. This auxiliary loss ensures that the prediction is internally consistent, and is what causes the clustering of the prediction. Note that if the model predicts $\rho = 1$ for one position and $0$ everywhere else, this metric would be $0$. As such, it encourages a tight spatial weight distribution, which is useful for separating closely packed objects. However, the dropout in the weight during training ensures that it does not collapse to a single pixel. Without this loss, LodeSTAR would occasionally ignore the feature maps, and instead tune the weights such that only the average is correct.

In all cases, LodeSTAR was trained using the Adam optimizer \cite{kingma2017adam}, with a learning rate of 0.001. Unless otherwise stated, the model was trained using $5000$ mini-batches of $8$ samples. The exception is for the 3D tracking described in Figure~3, where the model is trained on $15000$ mini-batches.

\subsection*{Particle detection criteria \label{ch:det}}

Particles are detected by finding local maxima in a score map using the function \texttt{h\_maxima} of the Python module \texttt{skimage}. The score map is based on two separate metrics. The first is the weight map $w$. The second is a clustering metric calculated as
\[
c_{ij}^{-1} = \sum_{k=0}^{K} \left(\sum_{j^\prime = j - 1}^{j+1} \sum_{i^\prime=i-1}^{i+1} \frac{P_{i^\prime j^\prime k}^2}{9^2}\right) - \left(\sum_{j^\prime = j - 1}^{j+1} \sum_{i^\prime=i-1}^{i+1} \frac{P_{i^\prime j^\prime k}}{9}\right)^2,
\]
where $P$ is the feature map produced by the network. This can be likened to convolving the feature map with a variance kernel. The two metrics are combined geometrically as $w^\alpha c^\beta$, where $\alpha$ and $\beta$ can be tuned to optimize performance. In the cases presented in this paper, we consistently use $\beta = 1 - \alpha$. For the cases in the chapter ``Validation with experimental data'', we use, in order, $alpha=0.2$, $alpha=1$, $alpha=1$, $alpha=1$. For the remaining cases, we use $alpha=0.1$ Local maxima in this product are taken as observation if they are larger than some threshold, which in turn is chosen as a quantile of all scores. The quantile can be chosen a priori based on the density of the experimental data, but it can commonly be taken as the $99^{\rm th}$ percentile.

\subsection*{Plankton preparation and imaging}

\emph{Noctiluca scintillans} (SWE2020) was isolated from the Swedish west coast in November 2020. The culture was maintained in $16^\circ{\rm C}$,  $26\,{\rm psu}$, and 12:12~h light:dark cycles. The culture flasks were shaded by a screen to limit growth of the food organisms, \emph{Dunaliella tertiolecta}. \emph{Noctiluca} cultures were fed \emph{ad libitum} 2-4 times per month. 
The planktons are imaged with an inline holographic microscope, illuminated with a LED source (Thorlabs M625L3) of center wavelength $632\,\rm{nm}$ (details in \cite{bachimanchi2022microplankton}). The images are recorded with a CMOS sensor (Thorlabs DCC1645C) placed at a distance of $\approx 1.5\,{\rm mm}$ from the sample well, at 10 frames per second and with an exposure time of $8\,{\rm ms}$.

\subsection*{Human neuroblastoma cell sample preparation}

SH-SY5Y cells were grown in cell culture media (CCM) containing a 1:1 mixture of minimal essential medium (MEM) and nutrient mixture F-12 Ham supplemented with 10\% heat-inactivated fetal bovine serum, 1\% MEM nonessential amino acids, and 2 mM l-glutamine. The cells were detached (trypsin-EDTA 0.25\%, 5 min) and passaged twice a week. The cells were tested and verified mycoplasma-free. Cells were plated 1 day prior to experiments in glass-bottomed culture dishes (MatTek; 25000 cells/14 mm glass region) for microscopy. 
Cells were washed 1x with serum-free CCM before exposure to $225\,{\rm nm}$ green fluorescent polystyrene particles diluted in serum-containing CCM with the addition of 1\% Penicillin-Streptomycin. After a $4$ hour incubation at $37^\circ{\rm C}$ 5\% CO2, the cells were washed twice for 2 minutes with serum-free CCM, followed by addition of serum-containing CCM supplemented with 1\% Penicillin-Streptomycin and 30 mM HEPES to buffer the medium. The cells were imaged at $37^\circ{\rm C}$ using a OKOlab stage top incubator. 

\subsection*{Holographic imaging}

The used monodisperse particles are and $0.15\,{\rm \upmu m}$ (modal radius) polystyrene (Invitrogen) and $0.23\,{\rm \upmu m}$ (modal radius, NIST-certified standard deviation ±6.8 nm) polystyrene (Polysciences) (sizes verified using nanoparticle tracking analysis performed by NanoSight). Samples were imaged under flow in straight hydrophilized channels with a height of $20\,{\rm \upmu m}$ and a width of $800\,{\rm \upmu m}$ in chips made from Topas (COC, ChipShop). The images were captured using a off-axis holographic microscope \cite{Midtvedt2021}, using a $633\,{\rm nm}$ HeNe laser (Thorlabs) and a  Olympus 40× 1.3 NA oil objective. The interference pattern was collected using a CCD camera (AlliedVision, ProSilica GX1920), at a frame-rate of 30 frames per second, and an exposure time of $2\,{\rm ns}$ to $4\,{\rm ms}$.

For the intracellular data, a $40\times$, 0.95 NA Objective (Nikon, CFI Plan Apokromat) objective was used instead. The particles consisted of 225-nm-radius green-fluorescent polystyrene spheres (PS-FluoGreen, microparticles Gmbh). The particles were diluted 5000 times in cell media from the stock solution concentration of 2.5 wt\% to a concentration of $5.25\,{\rm \upmu g\,ml^{-1}}$. Further, to improve the data aquisition quality, the stage was oscillated at $1.1\,{\rm \upmu m}$ roughly every 4 seconds and offset pairs of frames were subtracted from each other to mitigate noise from reflections. The sample was imaged at 3 frames per second.

The fluorescence arm was illuminated using a $465\,{\rm nm}$ LED excitation (CoolLED), and imaged using a ORCA-Flash 4.0 V2.0 CMOS camera (Hamamatsu). We filtered the signal using a $491\,{\rm nm}$ dichromatic mirror (Chroma Technology Corporation), and separated the fluorescence channel from the holography channel using a long-pass dichroic mirror ($605\,{\rm nm}$), as well as a $525\pm 39\,{\rm nm}$ bandpass emission filter from Thorlabs. 

\subsection*{Measuring particle polarizability}

Particles were detected in each frame using a neural network trained by LodeSTAR using a single observation. The observations were subsequently traced over time by linear sum assignment, with a distance threshold of $1\,{\rm \upmu m}$. The polarizability was averaged over measurements from each detection in a trace. The LodeSTAR-measured polarizability was calibrated against a known population as $\alpha =3V\frac{n_p^2 - n_m^2}{n_p^2 + 2 n_m^2}$, based on the Clausius–Mossotti relation.
Further, for the intracellular data, each detection was compared to a fluorescence channel. If the detection was within $400\,{\rm nm}$ of a fluorescence detection, it was considered a polystyrene particle. If $>95\%$ of detections in a trace were linked to a fluorescence detection, the trace was considered a trace of a polystyrene particle. If $<5\%$ of detections in a trace were linked to a fluorescence detection, the trace was considered a trace of an intracellular particle.

%\bibliographystyle{ieeetr}
%\bibliography{references}

\begin{thebibliography}{10}

\bibitem{Manzo2015}
C.~Manzo and M.~F. Garcia-Parajo, ``A review of progress in single particle
  tracking: from methods to biophysical insights,'' {\em Reports on Progress in
  Physics}, vol.~78, p.~124601, oct 2015.

\bibitem{Midtvedt2021Quantitative}
B.~Midtvedt, S.~Helgadottir, A.~Argun, J.~Pineda, D.~Midtvedt, and G.~Volpe,
  ``Quantitative digital microscopy with deep learning,'' {\em Applied Physics
  Reviews}, vol.~8, no.~1, p.~011310, 2021.

\bibitem{Krishnamurthy2020}
D.~Krishnamurthy, H.~Li, F.~B. du~Rey, P.~Cambournac, A.~G. Larson, E.~Li, and
  M.~Prakash, ``Scale-free vertical tracking microscopy,'' {\em Nature
  Methods}, vol.~17, pp.~1040--1051, 2020.

\bibitem{Geerts1987}
H.~Geerts, M.~D. Brabander, R.~Nuydens, S.~Geuens, M.~Moeremans, J.~D. Mey, and
  P.~Hollenbeck, ``Nanovid tracking: a new automatic method for the study of
  mobility in living cells based on colloidal gold and video microscopy,'' {\em
  Biophysical Journal}, vol.~52, pp.~775--782, 11 1987.

\bibitem{Zahid2018}
M.~U. Zahid, L.~Ma, S.~J. Lim, and A.~M. Smith, ``Single quantum dot tracking
  reveals the impact of nanoparticle surface on intracellular state,'' {\em
  Nature Communications 2018 9:1}, vol.~9, pp.~1--11, 5 2018.

\bibitem{Liu2017}
M.~Liu, A.~J. Sternbach, D.~N. Basov, T.~Zanon-Willette, R.~Lefevre,
  R.~Metzdorff, al, H.~Miller, Z.~Zhou, J.~Shepherd, A.~J.~M. Wollman, and
  M.~C. Leake, ``Single-molecule techniques in biophysics: a review of the
  progress in methods and applications,'' {\em Reports on Progress in Physics},
  vol.~81, p.~024601, 2017.

\bibitem{Newby2018}
J.~M. Newby, A.~M. Schaefer, P.~T. Lee, M.~G. Forest, and S.~K. Lai,
  ``Convolutional neural networks automate detection for tracking of
  submicron-scale particles in 2d and 3d,'' {\em Proceedings of the National
  Academy of Sciences of the United States of America}, vol.~115,
  pp.~9026--9031, 2018.

\bibitem{Helgadottir2019}
S.~Helgadottir, A.~Argun, and G.~Volpe, ``Digital video microscopy enhanced by
  deep learning,'' {\em Optica}, vol.~6, pp.~506--513, 2019.

\bibitem{Speiser2021}
A.~Speiser, L.~R. Müller, P.~Hoess, U.~Matti, C.~J. Obara, W.~R. Legant,
  A.~Kreshuk, J.~H. Macke, J.~Ries, and S.~C. Turaga, ``Deep learning enables
  fast and dense single-molecule localization with high accuracy,'' {\em Nature
  Methods 2021 18:9}, vol.~18, pp.~1082--1090, 9 2021.

\bibitem{Midtvedt2021}
B.~Midtvedt, E.~Olsén, F.~Eklund, F.~Höök, C.~B. Adiels, G.~Volpe, and
  D.~Midtvedt, ``Fast and accurate nanoparticle characterization using
  deep-learning-enhanced off-axis holography,'' {\em ACS Nano}, vol.~15,
  pp.~2240--2250, 2 2021.

\bibitem{Ulman2017}
V.~Ulman, M.~Maška, K.~E. Magnusson, O.~Ronneberger, C.~Haubold, N.~Harder,
  P.~Matula, P.~Matula, D.~Svoboda, M.~Radojevic, I.~Smal, K.~Rohr, J.~Jaldén,
  H.~M. Blau, O.~Dzyubachyk, B.~Lelieveldt, P.~Xiao, Y.~Li, S.~Y. Cho, A.~C.
  Dufour, J.~C. Olivo-Marin, C.~C. Reyes-Aldasoro, J.~A. Solis-Lemus,
  R.~Bensch, T.~Brox, J.~Stegmaier, R.~Mikut, S.~Wolf, F.~A. Hamprecht,
  T.~Esteves, P.~Quelhas, Ömer Demirel, L.~Malmström, F.~Jug, P.~Tomancak,
  E.~Meijering, A.~Muñoz-Barrutia, M.~Kozubek, and C.~Ortiz-De-Solorzano, ``An
  objective comparison of cell tracking algorithms,'' {\em Nature methods},
  vol.~14, p.~1141, 12 2017.

\bibitem{Crocker1996}
J.~C. Crocker and D.~G. Grier, ``Methods of digital video microscopy for
  colloidal studies,'' {\em Journal of Colloid and Interface Science},
  vol.~179, pp.~298--310, 4 1996.

\bibitem{Parthasarathy2012}
R.~Parthasarathy, ``Rapid, accurate particle tracking by calculation of radial
  symmetry centers,'' {\em Nature Methods}, vol.~9, pp.~724--726, 7 2012.

\bibitem{Tahara2018}
T.~Tahara, X.~Quan, R.~Otani, Y.~Takaki, and O.~Matoba, ``{Digital holography
  and its multidimensional imaging applications: a review},'' {\em Microscopy},
  vol.~67, pp.~55--67, 02 2018.

\bibitem{sibarita2005deconvolution}
J.-B. Sibarita, ``Deconvolution microscopy,'' {\em Microscopy Techniques},
  pp.~201--243, 2005.

\bibitem{Vestergaard2014diffusion}
C.~L. Vestergaard, P.~C. Blainey, and H.~Flyvbjerg, ``Optimal estimation of
  diffusion coefficients from single-particle trajectories,'' {\em Phys. Rev.
  E}, vol.~89, p.~022726, Feb 2014.

\bibitem{Zangle2014}
T.~A. Zangle and M.~A. Teitell, ``Live-cell mass profiling: an emerging
  approach in quantitative biophysics,'' {\em Nature Methods 2014 11:12},
  vol.~11, pp.~1221--1228, 11 2014.

\bibitem{Altman2020}
L.~E. Altman and D.~G. Grier, ``Catch: Characterizing and tracking colloids
  holographically using deep neural networks,'' {\em The Journal of Physical
  Chemistry B}, vol.~124, pp.~1602--1610, 3 2020.

\bibitem{YangCyto}
S.-A. Yang, J.~Yoon, K.~Kim, and Y.~Park, ``Measurements of morphological and
  biophysical alterations in individual neuron cells associated with early
  neurotoxic effects in parkinson's disease,'' {\em Cytometry Part A}, vol.~91,
  no.~5, pp.~510--518, 2017.

\bibitem{Ashraf2020}
S.~Ashraf, A.~Hassan~Said, R.~Hartmann, M.-A. Assmann, N.~Feliu, P.~Lenz, and
  W.~J. Parak, ``Quantitative particle uptake by cells as analyzed by different
  methods,'' {\em Angewandte Chemie International Edition}, vol.~59, no.~14,
  pp.~5438--5453, 2020.

\bibitem{deeptrackgithub}
B.~Midtvedt, S.~Helgadottir, A.~Argun, J.~Pineda, D.~Midtvedt, and G.~Volpe,
  ``Deeptrack-2.0.'' \url{https://github.com/softmatterlab/DeepTrack-2.0},
  2020.

\bibitem{caron2021emerging}
M.~Caron, H.~Touvron, I.~Misra, H.~Jégou, J.~Mairal, P.~Bojanowski, and
  A.~Joulin, ``Emerging properties in self-supervised vision transformers,''
  2021.

\bibitem{kingma2017adam}
D.~P. Kingma and J.~Ba, ``Adam: A method for stochastic optimization,'' 2017.

\bibitem{bachimanchi2022microplankton}
H.~Bachimanchi, B.~Midtvedt, D.~Midtvedt, E.~Selander, and G.~Volpe,
  ``Microplankton life histories revealed by holographic microscopy and deep
  learning,'' 2022.

\end{thebibliography}

\section{Code availability}

All source code and examples are made publicly available at the DeepTrack-2.1 GitHub repository\cite{deeptrackgithub}.

\section{Competing interests}

The authors declare no competing interests.

\section{Author contributions}

BM and GV conceived the method. BM designed, implemented and tested the method. 
JP, FS, DM and GV contributed to the development of the method.
HB and ES collected and imaged the plankton data. 
EO collected the holographic data, using an experimental setup and software developed by EO, BM, DM, and FH. 
EW and EKE provided the SH-SY5Y cell cultures and analyzed the polystyrene particle uptake.
EO performed the cell measurements, within a methodological framework designed by EO, FH, DM, EW, and EKE, and experimentally realized by EO.
GV supervised the work. 
BM, JP and GV drafted the paper and the illustrations. 
All authors revised the paper. 

\section{Acknowledgments}

The authors would like to acknowledge funweding from the H2020 European Research Council (ERC) Starting Grant ComplexSwimmers (Grant No. 677511), the Horizon Europe ERC Consolidator Grant MAPEI (Grant No. 101001267), the Knut and Alice Wallenberg Foundation (Grant No. 2019.0079), and Vetenskapsrådet (grant numbers 2016-03523 and 2019-05071).

\newpage

\section*{Supplementary Videos}

\subsection*{Video 1}
LodeSTAR successfully tracks particles of various shapes, orientations and noise.

\subsection*{Video 2}
Mouse hematopoietic stem cells are tracked by LodeSTAR as they divide. The score map used for detection is also shown.

\subsection*{Video 3}
Human hepatocarcinoma-derived cells are tracked by LodeSTAR. The score map used for detection is also shown.

\subsection*{Video 4}
Pancreatic stem cells are tracked by LodeSTAR as they divide. The score map used for detection is also shown.

\subsection*{Video 5}
\emph{Noctiluca scintillans} are tracked by LodeSTAR. LodeSTAR detects the optically dense area near the ventral groove where the nucleus is located. The score map used for detection is also shown.

\subsection*{Video 6}
\emph{Noctiluca scintillans} are tracked by LodeSTAR. LodeSTAR detects the plankton as a whole. The score map used for detection is also shown.

\subsection*{Video 7}
Polystyrene particles under flow are tracked by LodeSTAR in 3D.

\subsection*{Video 8}
SH-SY5Y cells with polysyterene particles inside and around the cell. LodeSTAR detects both the polystyrene (orange markers) and intracellular particles (blue markers).

\end{document}